\documentclass[pra,aps,twocolumn,showpacs,10pt]{revtex4-1}
\usepackage{amsmath,amssymb,graphicx}

\begin{document}

\title{Underlying conservation and stability laws in nonlinear propagation of axicon-generated Bessel beams}

\author{Miguel A. Porras}
\affiliation{Grupo de Sistemas Complejos, Universidad Polit\'{e}cnica de Madrid, ETSI de Minas y Energ\'{i}a, Rios Rosas 21, 28003 Madrid, Spain}
\author{Carlos Ruiz-Jim\'enez}
\affiliation{Grupo de Sistemas Complejos, Universidad Polit\'ecnica de Madrid, ETSI Agr\'onomos, Ciudad Universitaria s/n, 28040 Madrid, Spain}
\author{Juan Carlos Losada}
\affiliation{Grupo de Sistemas Complejos, Universidad Polit\'ecnica de Madrid, ETSI Agr\'onomos, Ciudad Universitaria s/n, 28040 Madrid, Spain}

\begin{abstract}
In light filamentation induced by axicon-generated, powerful Bessel beams, the spatial propagation dynamics in the nonlinear medium determines the geometry of the filament channel and hence its potential applications. We show that the observed steady and unsteady Bessel beam propagation regimes can be understood in a unified way from the existence of an attractor and its stability properties. The attractor is identified as the nonlinear unbalanced Bessel beam (NL-UBB) whose inward H\"ankel beam amplitude equals the amplitude of the linear Bessel beam that the axicon would generate in linear propagation. A simple analytical formula that determines de NL-UBB attractor is given. Steady or unsteady propagation depends on whether the attracting NL-UBB has a small, exponentially growing, unstable mode. In case of unsteady propagation, periodic, quasi-periodic or chaotic dynamics after the axicon reproduces similar dynamics after the development of the small unstable
mode into the large perturbation regime.
\end{abstract}

\maketitle

\section{Introduction}

The nonlinear propagation of very intense pulsed Bessel beams (BBs) has attracted a lot of attention in recent years, specially because of the ability of BBs of creating filamentary ionized channels that may be longer and more spatially controllable \cite{POLESANAPRA2008,XIE,JUKNA} that the filaments created by focusing standard, Gaussian-like light pulses \cite{BRAUN,COUAIRON0}. The versatility of Bessel beams for filamentation has been dramatically demonstrated very recently with the generation of tubular plasma channels when the seeding Bessel beam carries an optical vortex \cite{XIE,JUKNA}. These achievements have opened new perspectives in ultrafast laser material processing in transparent dielectrics, such as waveguide writing and micro- or nanomachining \cite{MARCINKEVICIUS,ARNOLD}, or in long-range filamentation in gases with application, for instance, in microwave guiding by filaments in the atmosphere \cite{CHATEAUNEUF}.

The first studies on BB propagation in nonlinear media date from the beginning of the past decade \cite{GADONAS,PYRAGAITE}. Nonlinear Bessel beams as stationary (non-diffracting) solutions to the nonlinear Schr\"odinger equation (NLSE) with Kerr nonlinearity were first introduced in \cite{JOHANNISSON}. Nonlinear unbalanced Bessel beams (NL-UBBs) \cite{PORRAS1} were later found as stationary solutions of the NLSE in media with Kerr nonlinearity and nonlinear losses (NLLs), just the two key nonlinearities determining the spatial dynamics in BB filamentation. These NL-UBBs have indeed been proven to play a prominent role in the filamentation with axicon focused BBs \cite{POLESANAPRA2008,COUAIRON,POLESANAPRL2007}, acting as attractors of the dynamics. Matter waves of this kind can also exist in Bose-Einstein condensates \cite{VICTOR}. More recently NL-UBBs carrying vortices have been also described \cite{PORRAS2}, and have similarly found to act as attractors in the filamentation seeded by axicon focused vortex BBs \cite{XIE,JUKNA}.

In the experimental and numerical studies on nonlinear BB propagation, two different initial conditions for the light entering the medium are usually considered. In a first arrangement, BBs are launched into the medium when they are already formed \cite{GADONAS,PYRAGAITE,POLESANAOPTEX2005,POLESANAPRE2006,POLESANAPRL2007,PORRAS2,GAIZAUKAS}, e. g., an ideal BB at any transversal plane, or an apodized BB at the focus of an axicon. Except if NLLs dominate initially the dynamics \cite{POLESANAOPTEX2005,PORRAS2,POLESANAPRE2006}, Kerr nonlinearity induces in this case large temporal and spatial instabilities \cite{POLESANAPRL2007,GAIZAUKAS}. In most of filamentation experiments with BBs \cite{POLESANAPRA2008,XIE,JUKNA,COUAIRON,POLESANAPRL2007} and related numerical studies \cite{ROSKEY}, the radiation exiting from the BB generator enters the nonlinear medium prior to the formation of the BB, in a state of widespread energy at low intensity levels, so that the linear BB is never formed. With an axicon, for example, the medium is placed in contact with it, or simply fills the space surrounding it, as in filamentation in gases. This ``soft" input condition has been proven useful to prevent the onset of large temporal instabilities in the nonlinear medium \cite{POLESANAPRL2007}. With this arrangement, two different Bessel beam propagation regimes have been observed \cite{POLESANAPRA2008,COUAIRON}. In a steady Bessel propagation regime, the input radiation undergoes a transformation into a quasi-stationary state within the Bessel zone that has been identified as a NL-UBB. In a unsteady regime, the light intensity and fluence feature periodic, quasi-periodic, even disordered spikes in the Bessel zone (and azimuthal breaking in the case of vortex BBs \cite{JUKNA}). This regime has been associated with sufficiently small cone angles and relatively low input powers so that self-focusing is the dominant nonlinearity.

In this paper we aim at providing a unified understanding of these two regimes of BB propagation under soft input conditions.
We show that these two regimes are different manifestations of the same underlying dynamics. Either steady or unsteady, the spatial dynamics is dominated by the existence of an attractor in the form of a specific NL-UBB. We identify the attracting NL-UBB and derive an approximate analytical expression specifying it in terms of the properties of the nonlinear medium and the light beam illuminating the axicon. However, an attractor is not necessarily a stable attractor; its instability may lead to a richer dynamics around it, including periodic, quasi-periodic and chaotic behavior. The unsteady or steady regimes are seen to be determined by the existence or not of a small, exponentially growing unstable mode of the attracting NL-UBB. In the unsteady regime, the unstable dynamics in the Bessel zone of the axicon is triggered by the unstable mode and is seen to reproduce its characteristic oscillation frequency, its development into large periodic or quasi-periodic anharmonic oscillations, or its development into chaotic oscillations, depending on the gain of the small unstable mode. Although the unsteady Bessel filamentation regime has been previously suggested to be associated with NL-UBB instability \cite{POLESANAPRL2007,COUAIRON}, it is only the identification of the attracting NL-UBB that have allowed us to analyze its stability properties, and hence to verify that hypothesis, putting it in quantitative terms.

For simplicity we focus on BBs generated by axicons in most of the numerical simulations, but the same results are seen to hold for other soft input conditions that would generated BBs in linear propagation. We illustrate the results in air at 800 nm, in which case the characteristic angles separating the different regimes are quite small, but we have verified that the same results hold at larger angles (but still paraxial) in condensed media.

\section{Nonlinear unbalanced Bessel beams}\label{NL-UBB}

We consider diffraction, Kerr nonlinearity and NLLs as the key effects determining the propagation of the light beam coming from the BB generator.
In the paraxial approximation, the envelope $A$ of the light beam $E=A\exp(-i\omega t + i k z)$ of frequency $\omega$ and propagation constant $k$, is then suitably described by the NLSE
\begin{equation}\label{NLSE}
\partial_z A = \frac{i}{2k}\Delta_\perp A + \frac{ik n_2}{n}|A|^2 A -\frac{\beta^{(M)}}{2}|A|^{2M-2}A \, ,
\end{equation}
where $n$, $n_2$ and $\beta^{(M)}$ are, respectively, the linear and nonlinear refractive indexes and the $M$-photon absorption coefficient. For the initial conditions of interest, and according to \cite{POLESANAPRL2007}, temporal effects are assumed to play a secondary role.
\begin{center}
\begin{figure}
\includegraphics[width=4.3cm]{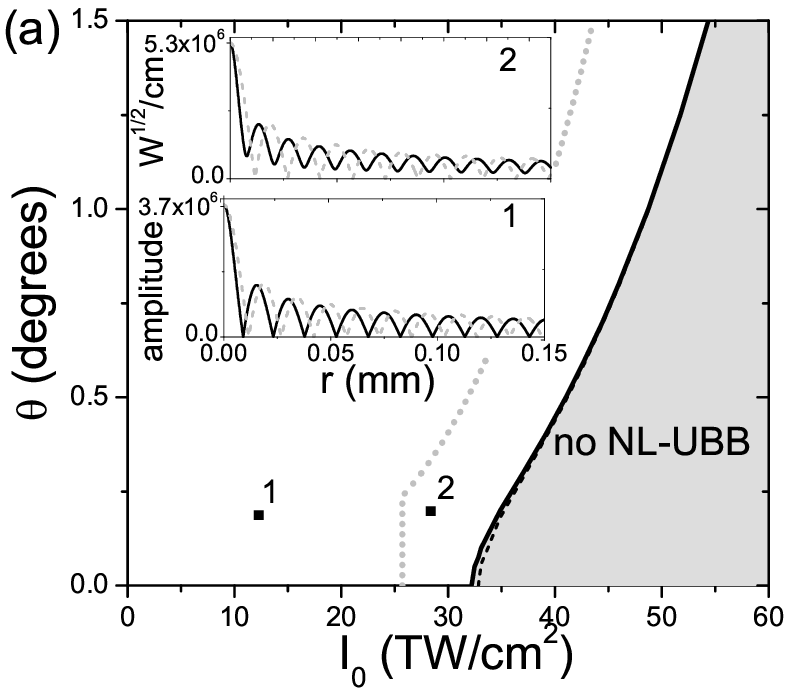}\includegraphics[width=4.4cm]{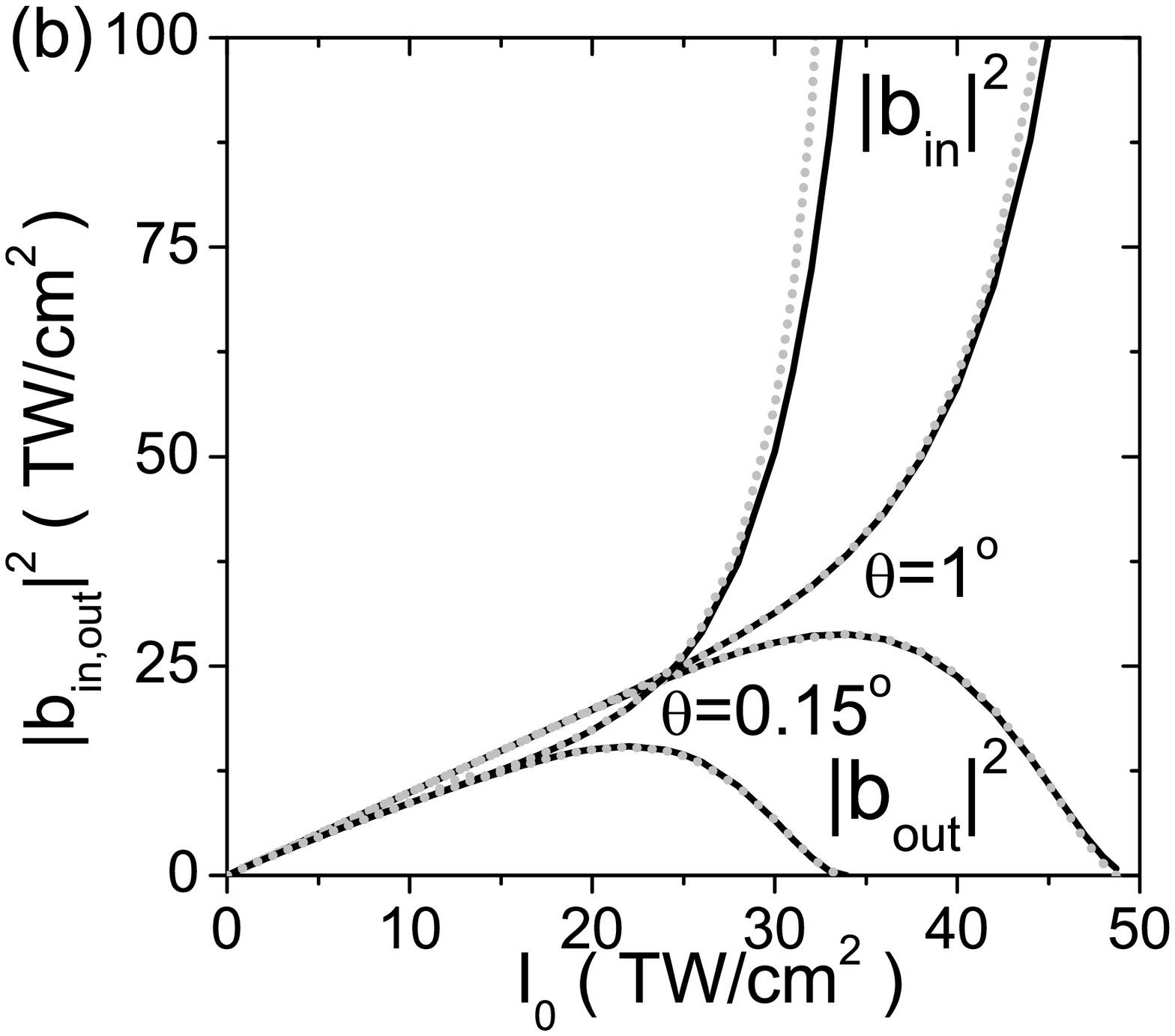}
\caption{\label{Fig1}(a) In air at 800 nm ($n\simeq 1$, $n_2=3.2\times 10^{-19}$ cm$^2$/W, $M=8$, $\beta^{(8)}=1.8\times 10^{-94}$ cm$^{13}$/W$^7$), region in parameters space ($I_0$,$\theta$) of existence of NL-UBBs. The dashed black curve is the approximate analytical curve ($I_{0,\rm max}$,$\theta$) given in the text. To the right of the gray dotted curve, NL-UBB are said to be NLL-dominated in the sense that $L_{\rm MPA}< L_{\rm dif}$ and $L_{\rm MPA}< L_{\rm Kerr}$, where the characteristic lengths are $L_{\rm MPA}=2/\beta^{(M)}I^{M-1}$ for multiphoton absortion, $L_{\rm dif}=1/|\delta|$ (Rayleigh range of central lobe of Bessel function) for diffraction, and $L_{\rm Kerr}=n/kn_2I_0$ for Kerr nonlinearity. Insets 1 and 2: Amplitude radial profiles of NL-UBBs with $\theta=0.15$ deg, and $I_0=12$ and $28$ TW/cm$^2$ (solid curves), and of linear BBs of the same intensity (dashed curves). The squares 1 and 2 indicate the location of those NL-UBBs in parameters space. (b) Values of $|b_{\rm in}|^2$ and $|b_{\rm out}|^2$ for NL-UBBs with $\theta=0.15$ and $1$ degrees, extracted from the numerically evaluated radial profiles (gray dotted curves), and approximately evaluated from Eq. (\ref{binout}) (solid curves).}
\end{figure}
\end{center}
In order to properly understand the propagation, it is important to review the properties of NL-UBBs, stressing their asymptotic properties. NL-UBBs were introduced in \cite{PORRAS1} as non-diffracting and non-attenuating solutions of (\ref{NLSE}) of the form $A=a(r)\exp[i\phi(r)]\exp(i\delta z)$, where $\delta=-k\theta^2/2$ is a negative axial wave vector shift corresponding (in the paraxial approximation) to a cone angle $\theta$. The real amplitude $a(r)>0$ and phase $\phi(r)$ satisfy
\begin{eqnarray}
a^{\prime\prime} + \frac{a'}{r} + k^2\theta^2 a - (\phi')^2 a + \frac{2k^2n_2}{n} a^3=0 \,, \label{a}\\
-F_r\equiv -\frac{1}{k} 2\pi r \phi' a^2 = \beta^{(M)} 2\pi\int_0^r dr r a^{2M} \equiv N_r \,, \label{phi}
\end{eqnarray}
(prime signs stand for $d/dr$) with boundary conditions $a(0)=\sqrt{I_0}$, $a'(0)=0$, $\phi'(0)=0$, where $I_0$ is the peak intensity of the NL-UBB. Equation (\ref{phi}) is the refilling condition for stationarity with nonlinear absorption, stating that the nonlinear power losses $N_r$ in each circle of radius $r$ are compensated by an inward radial flux $-F_r$ through its circumference. For each cone angle $\theta$, NL-UBB exist up to a maximum value $I_{0,\rm max}$ of the peak intensity whose value depends on the optical properties of the medium at $\omega$. Figure \ref{Fig1}(a) shows the region of existence in the parameters space ($I_0$,$\theta$) of NL-UBBs in air at 800 nm, and the insets two typical radial amplitude profiles. An approximate formula relating $I_{0,\max}$ to $\theta$ is $\theta^2=\sigma_M\beta^{(M)}I_{0,\rm max}^{M-1}/k_0- 2n_2 I_{0,\rm max}/n_0$ [dashed curve in Fig. \ref{Fig1}(a)], where $\sigma_M=0.542,0.381,0.313,0.295,0.244,0.223\dots$ for $M=3,4,\dots$ \cite{PORRAS1}. At large radius $r$, NL-UBBs behave asymptotically as BBs, but with unbalanced amplitudes of its outward and inward H\"ankel components:
\begin{equation}\label{ASYMP}
a(r)e^{i\phi(r)}\simeq \frac{1}{2}\left[b_{\rm out} H_0^{(1)}(k\theta r)+
b_{\rm in} H_0^{(2)}(k\theta r)\right]\, ,
\end{equation}
with $|b_{\rm in}|\ge |b_{\rm out}|$, and only $b_{\rm in}=b_{\rm out}=\sqrt{I_B}$ for a linear BB of intensity $I_B$. The amplitudes $|b_{\rm in}|$ and $|b_{\rm out}|$ can be easily extracted from the radial intensity profile, known numerically or experimentally: Using the asymptotic forms of H\"ankel functions for large argument one gets, from Eq. (\ref{ASYMP}), $a^2\simeq (a_m^2/r)[1+C\cos(2k\theta r+\varphi)]$, with $C=2|b_{\rm out}||b_{\rm in}|/(|b_{\rm out}|^2 + |b_{\rm in}|^2)$ and $a^2_m=(|b_{\rm out}|^2+|b_{\rm in}|^2)/(2\pi)$, meaning that the asymptotic radial intensity profile consists of oscillations of contrast $C\le 1$ about an average value $a_m^2/r$. From these features, the H\"ankel amplitudes are obtained to be $|b_{\rm in, out}|^2=\pi a_m^2(1\pm\sqrt{1-C^2})$. Examples of their values extracted from the numerical intensity profiles are depicted in Fig. \ref{Fig1}(b) for increasing NL-UBB intensities $I_0$ at two cone angles.

\section{On the dynamics of real Bessel beams in nonlinear media}\label{DYNAMICS}

In Ref. \cite{POLESANAPRE2006}, it was demonstrated experimentally that a BB $A(r,0)=\sqrt{I_B}J_0(k\theta r)$ launched in a nonlinear medium in a regime where NLLs are significant transforms spontaneously into a NL-UBB that preserves the cone angle. In the ideal case of BBs carrying infinite power (and for the broader class of vortex NL-UBBs) the specific attracting NL-UBB has been identified as that whose inward H\"ankel amplitude equal the amplitude of the launched BB, that is, $|b_{\rm in}|=\sqrt{I_B}$ \cite{PORRAS2}. Since for the input BB $b_{\rm in}=b_{\rm out}=\sqrt{I_B}$, the amplitude of the inward H\"ankel component can be said to be a preserved quantity in the nonlinear dynamics.

In actual settings the input power is finite, and the medium is placed close to or in contact with an axicon (or other BB generators), or simply fills the space surrounding the axicon, so that the BB is not formed when the radiation enters the medium and propagates linearly initially. With an axicon, for instance, the field entering the medium at $z=0$ is usually modelled by
\begin{equation}
A(r,0)=\sqrt{I_G}\exp(-r^2/w^2)\exp(-ik \theta r),
\end{equation}
where $w$ and $I_G$ are the width and the intensity of the Gaussian beam illuminating the axicon. In linear propagation, this would produce an apodized BB of intensity $I_B= \pi k w \theta I_G / \sqrt{e}$ at a distance $z_B= w/2\theta$, which is one-half the length $w/\theta$ of the so-called Bessel zone. Under these soft input conditions, the unsteady Bessel filamentation regime, associated with small cone angles and relatively low intensities, results in periodic or quasi-periodic field oscillations \cite{POLESANAPRA2008,COUAIRON}, though it can also result in chaos (see below). The steady filamentation regime, associated with large cone angles or higher intensities, has been explained in terms of the formation of a NL-UBB \cite{POLESANAPRL2007,POLESANAPRA2008,COUAIRON}.

As pointed out in the introduction, these regimes are shown here to be different manifestations of the same underlying dynamics. Either steady or unsteady the spatial dynamics in BB filamentation is dominated by the existence of an attractor in the form of a specific NL-UBB, unsteady or steady regimes being determined by the existence or not of a small, exponentially growing unstable mode of the attracting NL-UBB. In Section \ref{ATTRACTOR} we identify the attractor and obtain approximate analytical formulas determining it. In Section \ref{INSTABILITY} we perform a linearized stability analysis of ideal NL-UBB and note a biunivocal relation between NLUBB stability/insability under small perturbations and steady/unsteady propagation after the real BB generator. The unsteady Bessel regime appears then to be triggered by the existence of a small unstable mode in the attracting NL-UBB. Depending on the gain, signatures of this mode, or of its development into large periodic, quasi-periodic or chaotic perturbation regimes are indeed observed in the Bessel zone of the axicon.

\section{The attracting nonlinear unbalanced Bessel beam}\label{ATTRACTOR}

We identify the attracting NL-UBB as that whose inward H\"ankel amplitude coincides with the amplitude of the BB that the BB generator would create in linear propagation, i. e., the NL-UBB with $|b_{\rm in}|=\sqrt{I_B}$. Since $b_{\rm in}=b_{\rm out}=\sqrt{I_B}$ for BBs, the amplitude of the inward H\"ankel component is not affected by nonlinearities, and in this sense can be said to be conserved. This conclusion is extracted from extensive numerical simulations, of which only a few examples are shown. Conceptually, it is not difficult to understand that the inward H\"ankel component created by the axicon, even if of finite power, and supplying power conically inwards is not affected by nonlinear absorption at the beam center in the Bessel zone.
\begin{center}
\begin{figure}
\includegraphics[width=4.3cm]{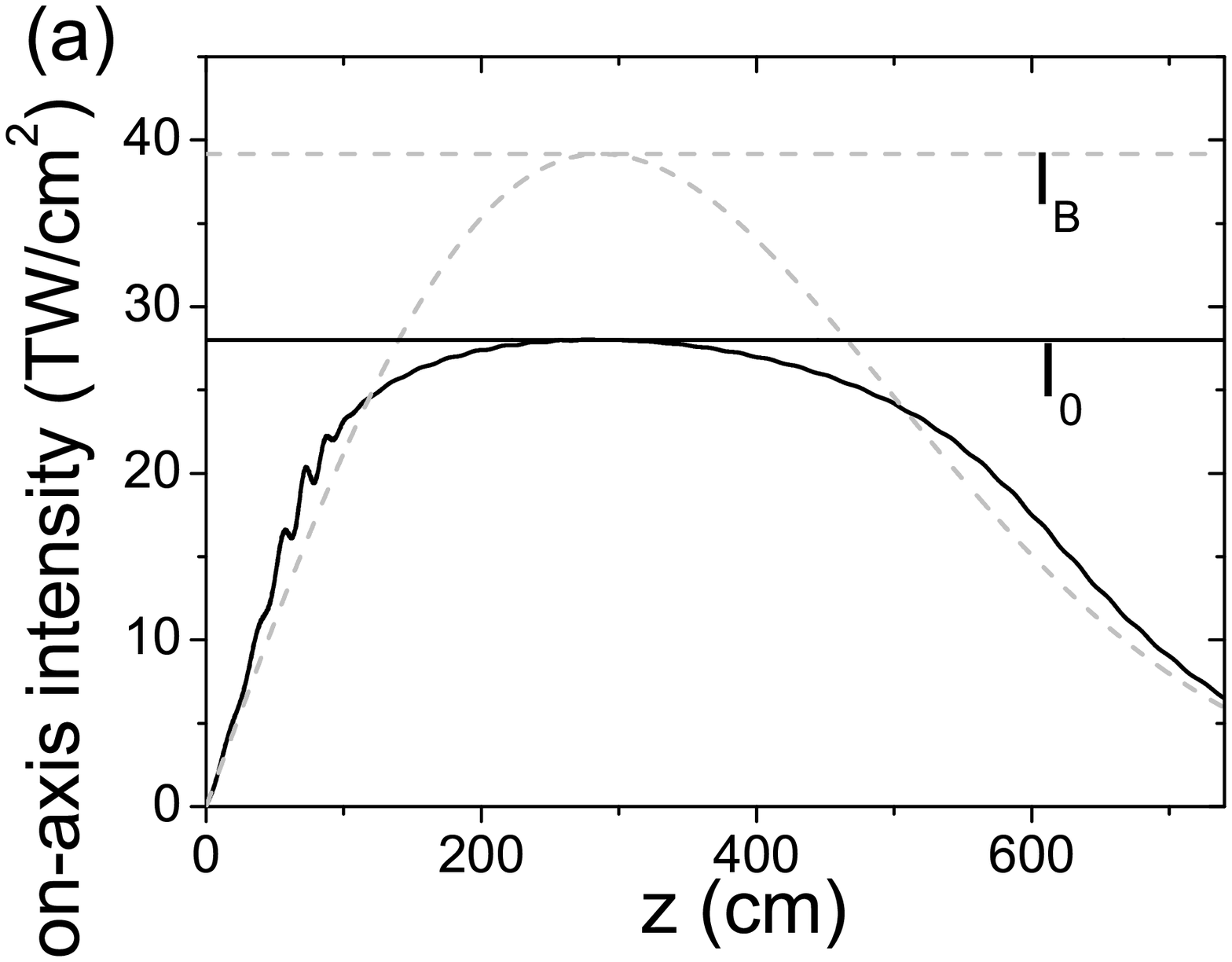}\includegraphics[width=4.3cm]{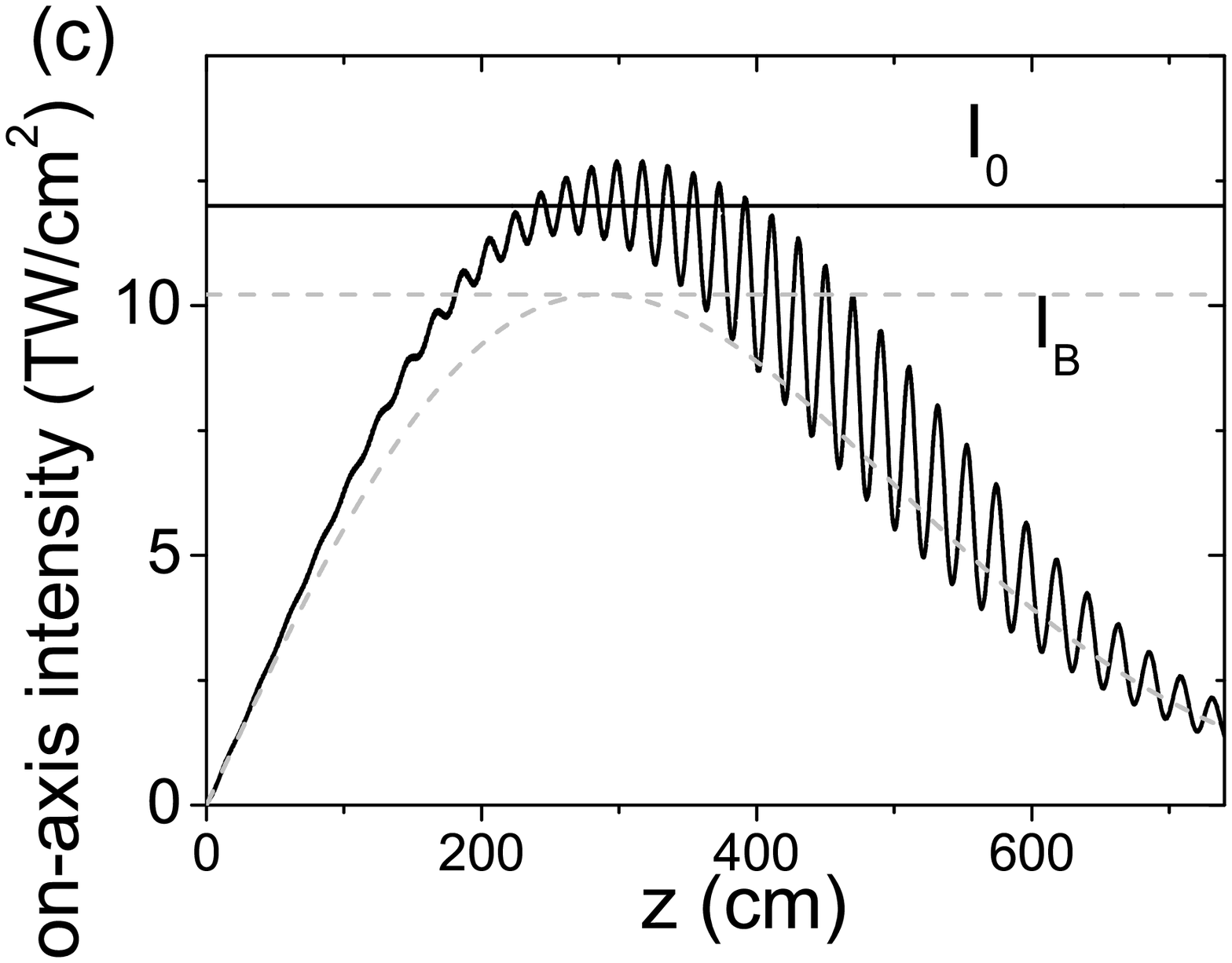}\\
\includegraphics[width=4.3cm]{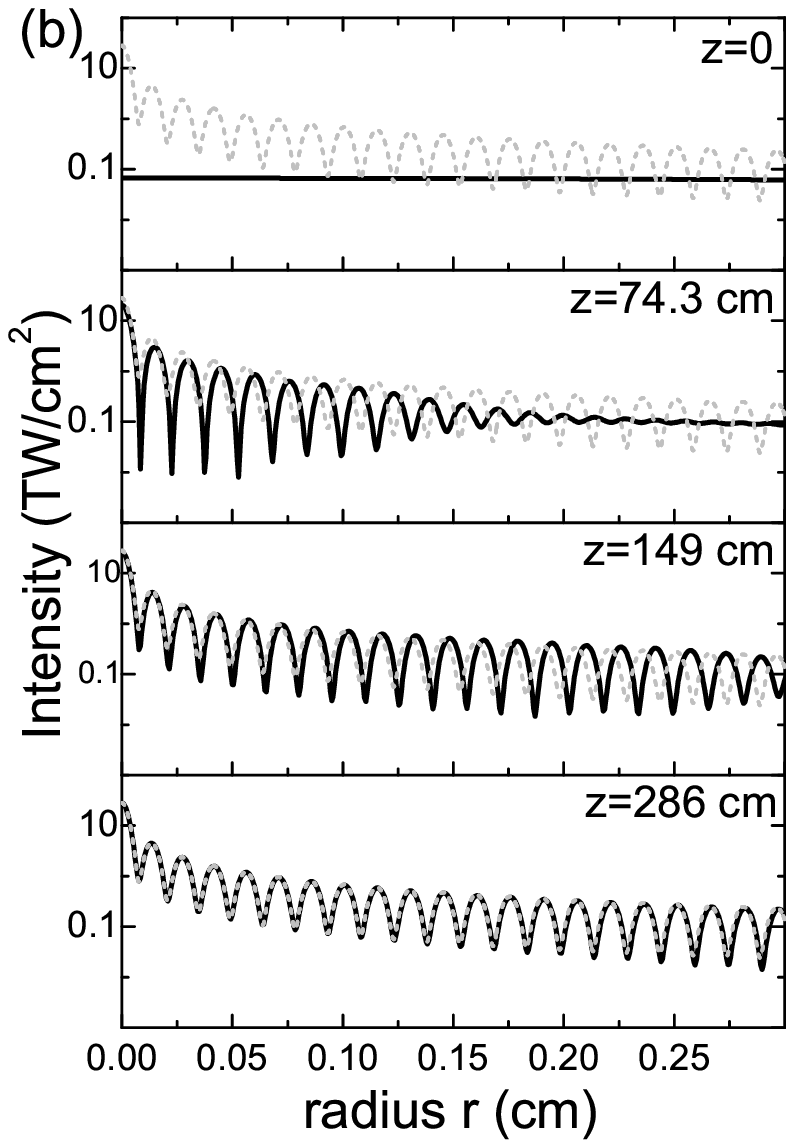}\includegraphics[width=4.3cm]{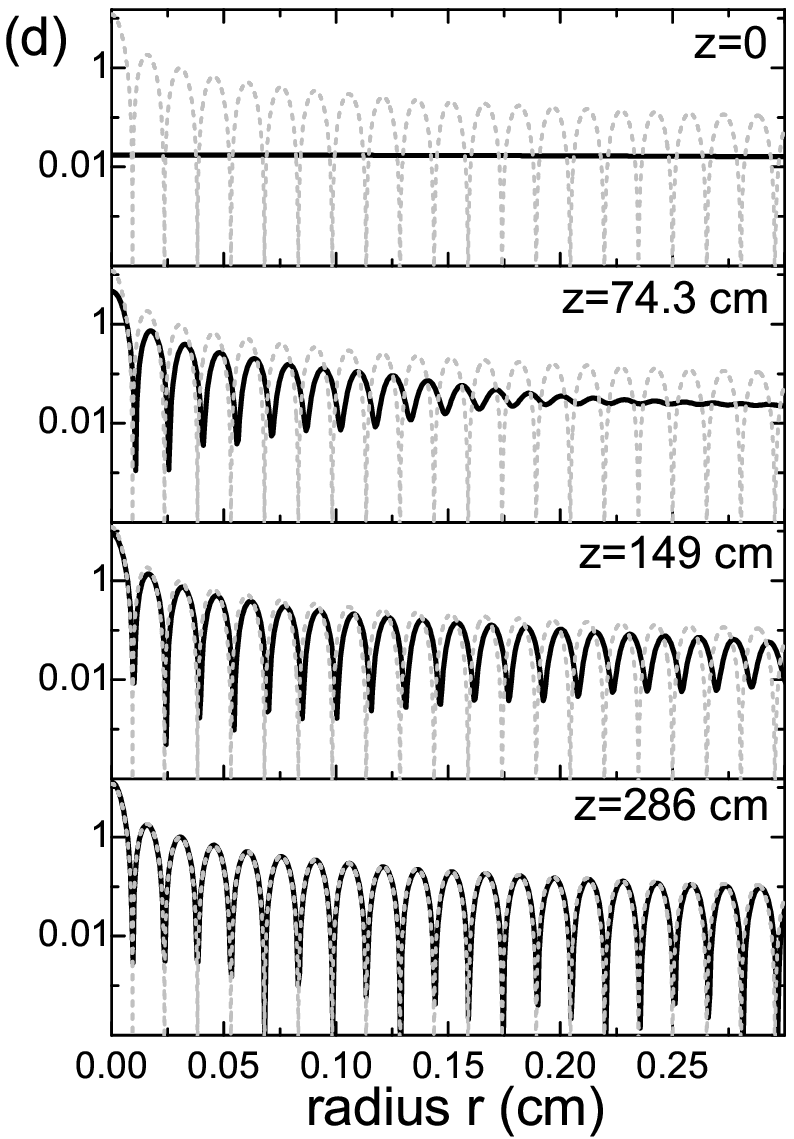}
\caption{\label{Fig2} (a) For a Gaussian beam of with $w=1.5$ cm and peak intensity $I_G=0.0666$ TW/cm$^2$ illuminating an axicon that forms a BB of cone angle $\theta=0.150$ deg and peak intensity $I_B=39.17$ TW/cm$^2$ at the center $z_B\simeq 286$ cm of the Bessel zone in linear propagation (dashed curve and horizontal dashed line), on-axis intensity in nonlinear propagation in air at 800 nm (solid curve), and numerically evaluated intensity $I_0=28$ TW/cm$2$ of the NL-UBB with $|b_{\rm in}|^2=I_B$ (horizontal solid line). (b) Radial intensity profile at increasing propagation distances up to the center of the Bessel zone (solid curves), and radial intensity profile of the attracting NL-UBB with $|b_{\rm in}|^2=I_B$ (gray dashed curve). (c) and (d) The same as in (a) and (b) but with $I_G=0.0174$ TW/cm$^2$ for the input Gaussian beam, $I_B=10.22$ TW/cm$^2$ for the linear BB, and $I_0=12$ TW/cm$^2$ for the NL-UBB with $|b_{\rm in}|^2=I_B$.}
\end{figure}
\end{center}
\begin{center}
\begin{figure}
\includegraphics[width=8.3cm]{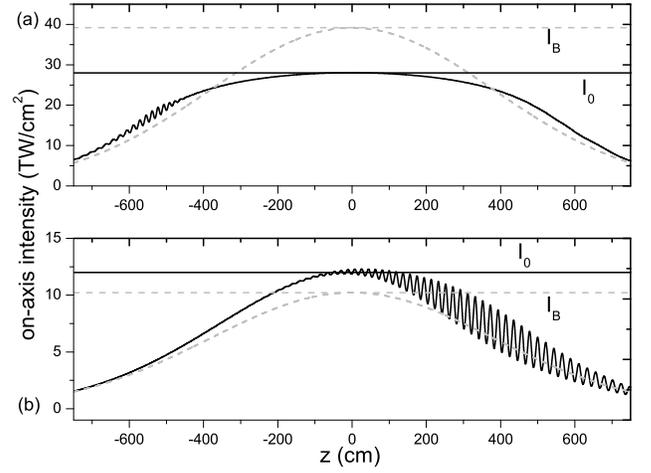}
\caption{\label{Fig3} The same as in Fig. \ref{Fig2} (a) and (c) but the beam entering into the medium is the Bessel-Gauss beam of Eq. (\ref{BG}) with $v=2$ cm at $z=-929$ cm from the waist at $z=0$. In (a), $I_B=39.17$ TW/cm$^2$ leading to $I_0=28$ TW/cm$^2$,  and in (b), $I_B=10.22$ TW/cm$^2$, leading to $I_0=12$ TW/cm$^2$.}
\end{figure}
\end{center}

Figure \ref{Fig2} illustrates this law for an axicon imprinting a cone angle $\theta=0.15$ deg illuminated by Gaussian beams of width $w=1.5$ cm and intensities $I_G=0.0666$ and $I_G=0.0174$ TW/cm$^2$ and propagating in air at $\lambda=800$ nm. In linear propagation these Gaussian beams would create linear BBs [dashed curves in Figs. \ref{Fig2}(a) and (c)] of intensities $I_B=39.17$ TW/cm$^2$ and $I_B=10.22$ TW/cm$^2$, respectively [dashed horizontal lines $I_B$ in Figs. \ref{Fig2}(a) and (c)]. In the steady regime of Fig. \ref{Fig2}(a), the on-axis intensity [solid curve in Fig. \ref{Fig2}(a)] in the Bessel zone stabilizes in the numerically evaluated intensity $I_0=28$ TW/cm$^2$ (solid horizontal line $I_0$) of the NL-UBB having $|b_{\rm in}|=\sqrt{I_B}=39.17$ [see Fig. \ref{Fig1}(b)]. The whole beam transforms in fact into the attracting NL-UBB, as seen in Fig. \ref{Fig2}(b) showing intensity profiles at increasing distances up to the center of the Bessel zone $z_B=286$ cm.
In the unsteady regime of Fig. \ref{Fig2}(c), the on-axis intensity [solid curve in Fig. \ref{Fig2}(c)] also approaches, but now oscillates about the numerically evaluated intensity $I_0=12$ TW/cm$^2$ (solid horizontal line $I_0$) of the NL-UBB with $|b_{\rm in}|=\sqrt{I_B}=10.22$ TW/cm$^2$ [see Fig. \ref{Fig1}(b)], and the same happens to the whole radial profile at increasing distances. Oscillations may be much more pronounced and disordered, but clear signatures of the attracting NL-UBB, its dominant small unstable mode, and its development into a large perturbation regime, are always observable, as shown below (see Fig. \ref{Fig7}).

The law $|b_{\rm in}|=\sqrt{I_B}$ holds for other finite-power versions of BBs, as the Bessel-Gauss beam
\begin{eqnarray}\label{BG}
A(r,z)&=& \sqrt{I_B}\frac{v^2}{v^2+\frac{2iz}{k}}J_0\left(\frac{v^2}{v^2+ \frac{2iz}{k}}k\theta r\right) \nonumber \\
&\times& \exp\left(-\frac{r^2+\frac{z^2\beta^2}{k^2}}{v^2+ \frac{2iz}{k}}\right)\, ,
\end{eqnarray}
producing at $z=0$ the Gaussian-apodized BB $\sqrt{I_B}J_0(k\theta r)e^{-r^2/v^2}$. For a soft input into the medium, the entrance plane is at $z=z_{\rm in}\ll 0$ such that the intensity is low enough for nonlinear effects to be initially negligible. For $v=2$ cm and $z_{\rm in}=-929$ cm,  propagation in air at the same wave length and the same linear BB peak intensities $I_B=39.17$ and $I_B=10.22$ TW/cm$^2$ as in Fig. \ref{Fig2}, the attracting NL-UBBs are seen in Fig. \ref{Fig3} to have the same intensities $I_0=28$ and $I_0=12$ TW/cm$^2$ as with the axicon.

Thus, given the intensity $I_B$ that a BB generator would create, it is possible to foresee the attracting NL-UBB. In practice, this requires to extract the values of $|b_{\rm in}|^2$ from the numerical radial profiles of NL-UBBs of different intensities $I_0$ with the given cone angle in the particular medium, as explained above [dotted curve $|b_{\rm in}|^2$ in Fig. \ref{Fig1}(b)] and in Ref. \cite{PORRAS2}, and to pick up the particular NL-UBB with $|b_{\rm in}|^2=I_B$. This long numerical procedure would be greatly simplified if we had analytical expressions for $|b_{\rm in}|^2$ as functions of the NL-UBB parameters ($\theta$, $I_0$) and the optical properties of the medium.

An approximate expression can be obtained as follows. We first note that Eq. (\ref{ASYMP}) implies $-F_\infty= (|b_{\rm in}|^2- |b_{\rm out}|^2)/k=N_{\infty}$ \cite{PORRAS1}, meaning that the unbalance of the H\"ankel amplitudes sets a net constant inward radial power flux coming from a reservoir at large radial distances to refill the total NLL $N_\infty$ during the propagation. At the same time, most of NLL take place in the beam center, where the NL-UBB can be approached, if the NL-UBBs is not well within the NLL-dominated region (see caption of Fig. \ref{Fig1}), by $A\simeq \sqrt{I_0}J_0(\sqrt{k^2\theta^2 + 2 k k_{\rm NL}}\,r)\exp(i\delta z)$, with $k_{\rm NL}= kn_2I_0/n$ \cite{PORRAS1}. Evaluation of $N_\infty$ with this profile yields the following approximate expression for the NLL of a NL-UBB, and hence an approximate relation between $|b_{\rm in}|$ and $|b_{\rm out}|$:
\begin{equation}\label{ONE}
k N_\infty = |b_{\rm in}|^2- |b_{\rm out}|^2 \simeq  \frac{\beta^{(M)}I_0^M}{k\theta^2(1 + 2n_2I_0/n\theta^2)} \gamma^{(M)}
\end{equation}
where $\gamma^{(M)}\equiv 2\pi\int_0^\infty J_0^{2M}(x)x dx$ is a number.

Second, numerical evaluation of $|b_{\rm in, out}|$ reveals that, except for NLL-dominated NL-UUBs (see caption of Fig. \ref{Fig1}), their average value $(|b_{\rm out}|+|b_{\rm in}|)/2$ can be approximated by the value $|b_{\rm in}|=|b_{\rm out}|\equiv |b_{\rm Kerr}|$ of the asymptotic form of the solutions of Eqs. (\ref{a}) and (\ref{phi}) in the absorption-less case (i. e., with $\beta^{(M)}=0$). Without absorption, $\phi(r)=0$ and the amplitude of nonlinear BBs behaves as $a(r)\simeq \frac{1}{2}[b_{\rm Kerr} H_0^{(1)}(k\theta r)+ b^{\star}_{\rm Kerr}H_0^{(2)}(k\theta r)]$ at large $r$. The scaling $\rho=k\theta r$ and $\tilde a =a/\sqrt{I_0}$ in Eqs. (\ref{a}), leads to the one-parameter problem $\tilde a^{\prime\prime}+ \tilde a^{\prime}/\rho + \eta \tilde a^3=0$, with initial conditions $\tilde a(0)=1$, $\tilde a'(0)=0$, and where $\eta=2n_2I_0/n\theta^2$. From the numerical solution of this problem with different values of $\eta$, we find the value of $|\tilde b_{\rm Kerr}|$ of the scaled asymptotic form $\tilde a(\rho)\simeq \frac{1}{2}[\tilde b_{\rm Kerr} H_0^{(1)}(\rho)+ \tilde b^{\star}_{\rm Kerr}H_0^{(2)}(\rho)]$ as a function of $\eta$, which is found to fit accurately to the function $|\tilde b_{\rm Kerr}|=f(\eta)=(1+c\eta)/(1+d\eta)$ with $c=0.63$ and $d=0.76$. Coming back to real variables, $|b_{\rm Kerr}|= \sqrt{I_0}(1+c\eta)/(1+d\eta)$ provides semi-analytical solution to the asymptotic form of nonlinear BBs in transparent media. Finally, since $(|b_{\rm out}|+|b_{\rm in}|)/2\simeq |b_{\rm Kerr}|$ in the nonlinearly lossy medium, we obtain
\begin{equation}\label{TWO}
|b_{\rm out}|+|b_{\rm in}| = 2 \frac{1+c(2n_2I_0/n\theta^2)}{1+d(2n_2I_0/n\theta^2)} \sqrt{I_0}\,.
\end{equation}
The two relations (\ref{ONE}) and (\ref{TWO}) lead to the approximate formulas
\begin{equation}\label{binout}
|b_{\rm in,out}|\simeq f(\eta)\sqrt{I_0} \pm \gamma^{(M)}\frac{\beta^{(M)}I_0^M}{4k\theta^2(1+\eta)f(\eta)\sqrt{I_0}}\, ,
\end{equation}
for the amplitudes of the inward and outward Hankel components of NL-UBBs as functions of their cone angle $\theta$ and peak intensity $I_0$ and the medium properties [see Fig. \ref{Fig1}(b), solid curves].

\begin{center}
\begin{figure}
\includegraphics[width=5.5cm]{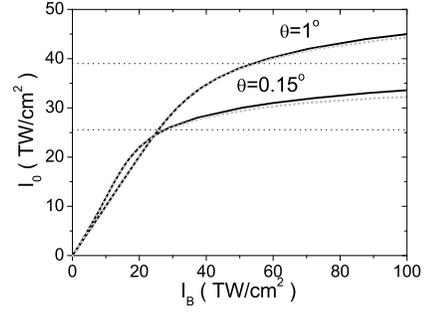}
\caption{\label{Fig4} In air at 800 nm and for the indicated cone angles, intensity $I_0$ of the attracting NL-UBB as a function of the intensity $I_B$ of the linear BB that the BB generator would create in linear propagation, numerically evaluated (dotted gray curves), and obtained from Eq. (\ref{I0}) (solid curves). Above the horizontal dotted lines NL-UBBs are NLL-dominated.}
\end{figure}
\end{center}

If we now set $|b_{\rm in}|= \sqrt{I_B}$ in Eq. (\ref{binout}), we obtain the approximate equation
\begin{equation}\label{I0}
\sqrt{I_B}\simeq f(\eta)\sqrt{I_0} + \gamma^{(M)}\frac{\beta^{(M)}I_0^M}{4k\theta^2(1+\eta)f(\eta)\sqrt{I_0}}
\end{equation}
($\eta=2n_2I_0/n\theta^2$) relating the intensity $I_0$ of the attracting NL-UBB to the cone angle $\theta$ and the intensity $I_B$ of the linear BB that the Bessel-beam generator would create. As an example, the values of $I_0$ provided by Eq. (\ref{I0}) for NL-UBBs in air at 800 nm, two cone angles $\theta$ and increasing $I_B$ are plotted in Fig. \ref{Fig4}, and are seen to match quite accurately the numerically obtained values for NL-UBBs of intensities $I_0$ below the NLL-dominated case (horizontal dotted lines). Equation (\ref{I0}) is seen to give a reasonably good estimate of $I_0$ even at huge intensities of $I_B=100$ TW/cm$^2$ well within the NLL-dominated case.

\begin{center}
\begin{figure}
\includegraphics[width=8.5cm]{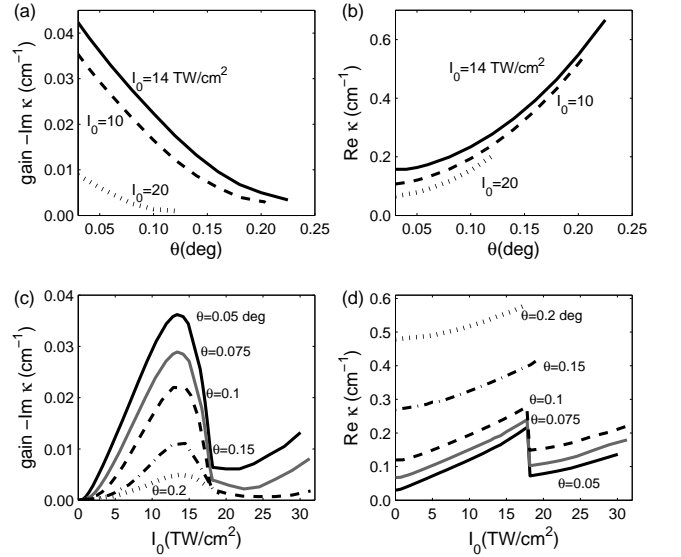}
\caption{\label{Fig5} (a) Gain $-\mbox{Im}\kappa$ and (b) oscillation frequency $\mbox{Re}\kappa$ of the most unstable mode mode of NL-UBBs in air at $800$ nm as functions of their cone angle for a few values of their peak intensity $I_0$. (c) and (d) The same as in (a) and (b) but as functions of the peak intensity $I_0$ for a few values of the cone angle $\theta$.}
\end{figure}
\end{center}

\section{Steady and unsteady regimes versus NL-UBB stability}\label{INSTABILITY}

Once the attractor is specified, our numerical simulations indicate that there is a bi-univocal relation between steady/unsteady propagation regime and stability/instability of the attracting NL-UBB against small radial perturbations.
\begin{figure*}
\begin{center}
\includegraphics[width=16cm]{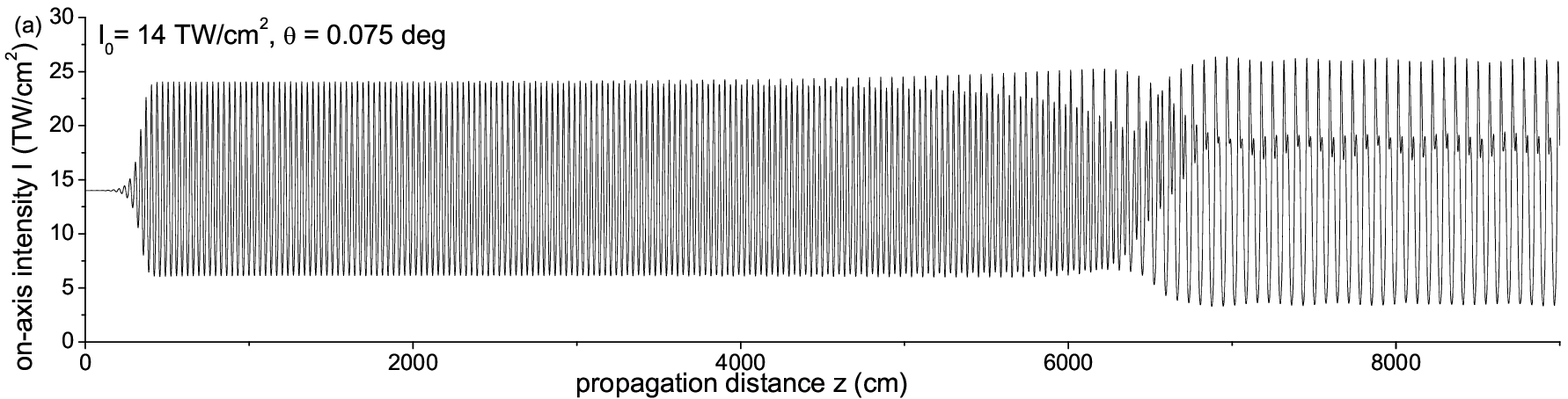}
\includegraphics[width=5cm]{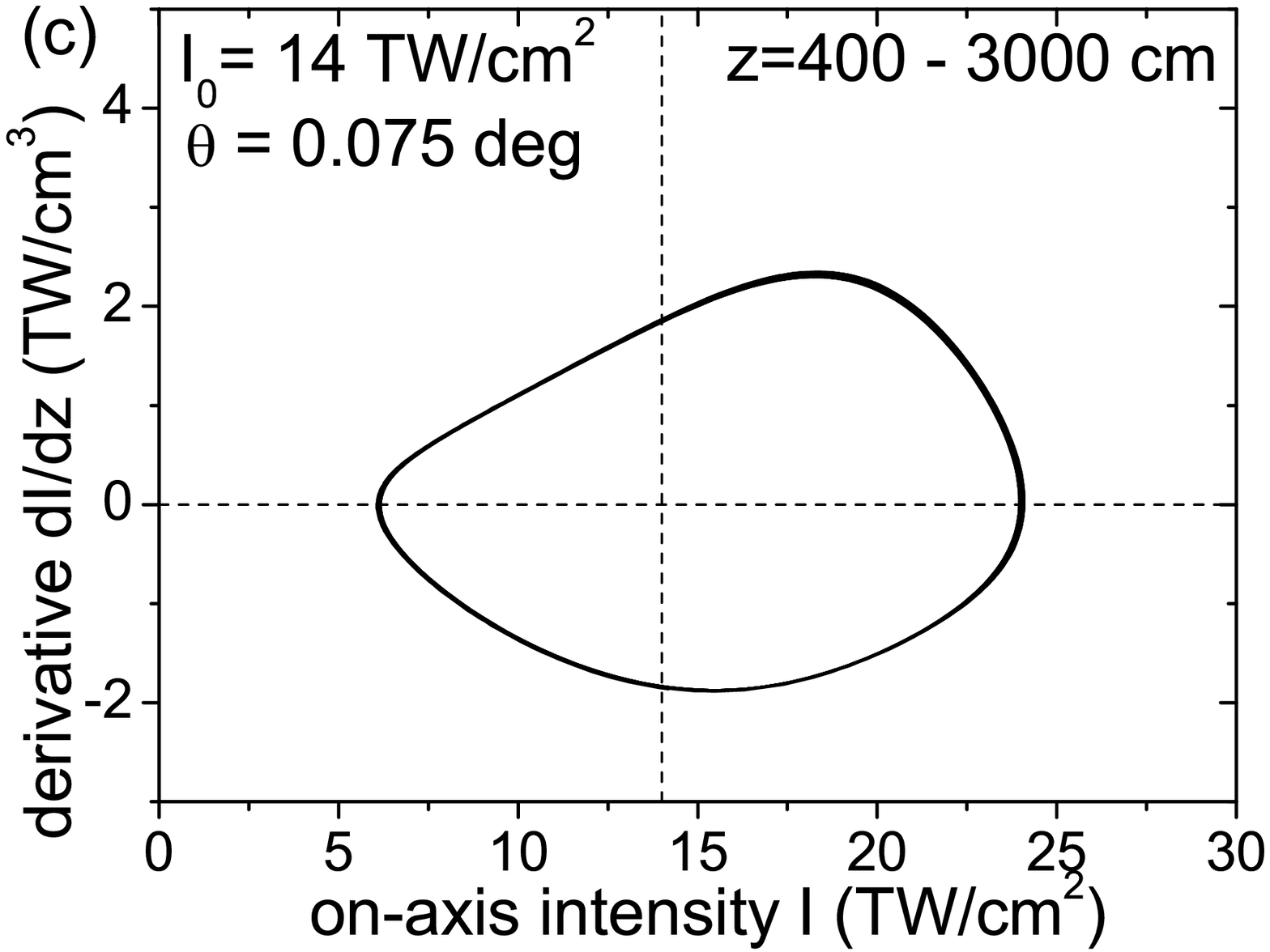} \hspace{1cm}\includegraphics[width=5cm]{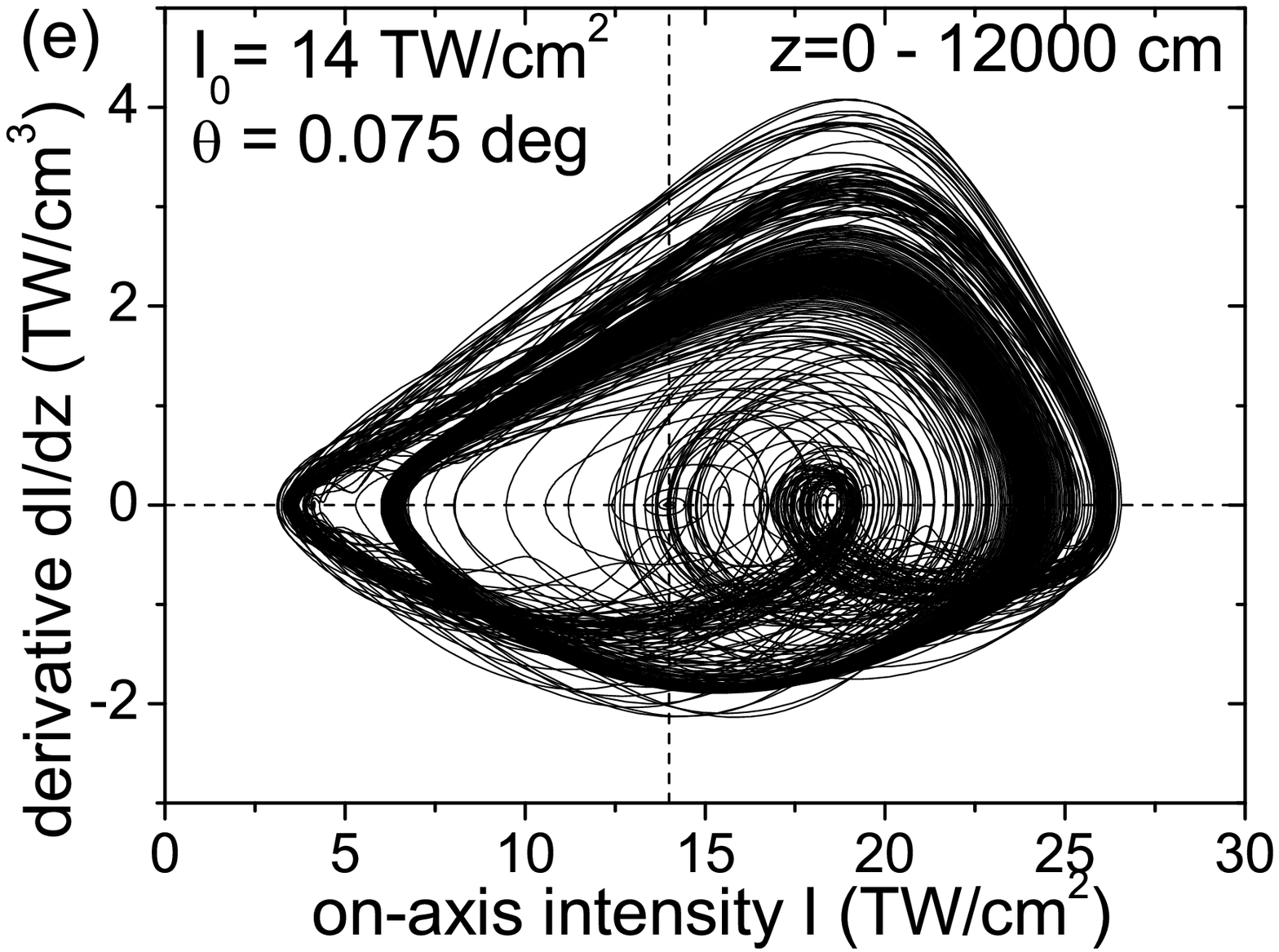}
\includegraphics[width=16cm]{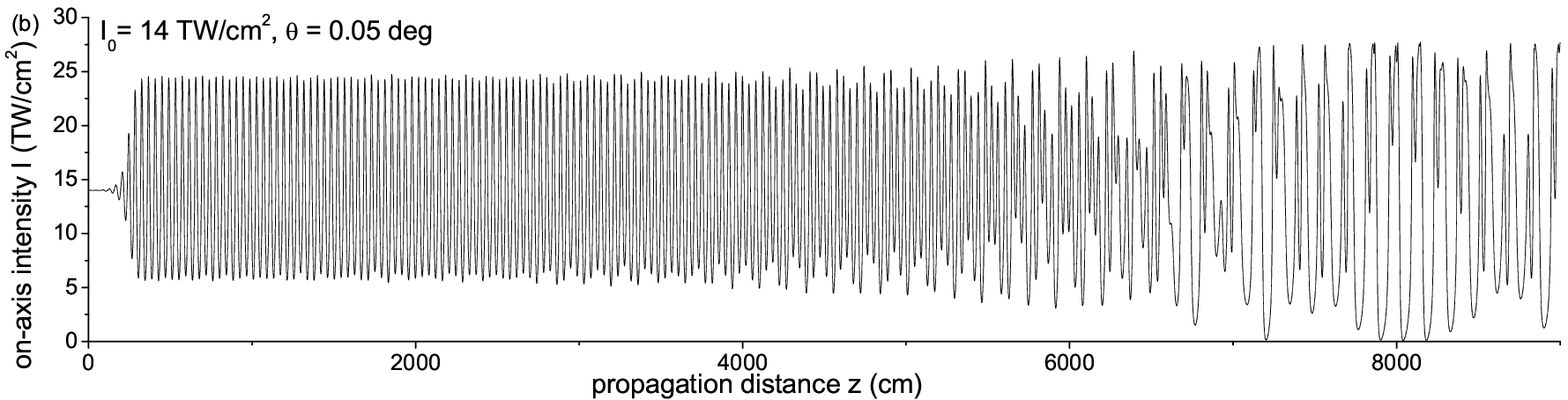}
\includegraphics[width=5cm]{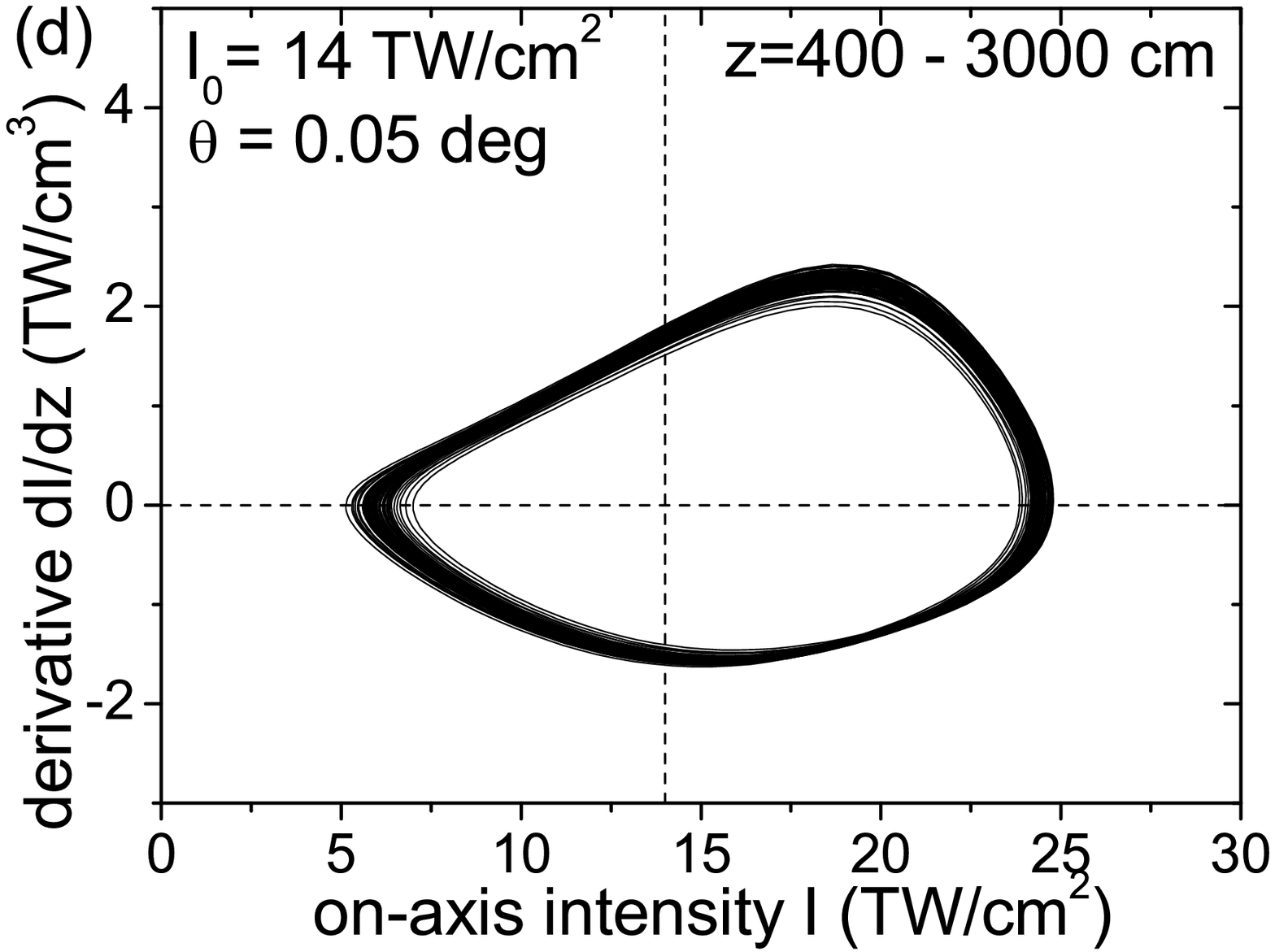}\hspace{1cm}\includegraphics[width=5cm]{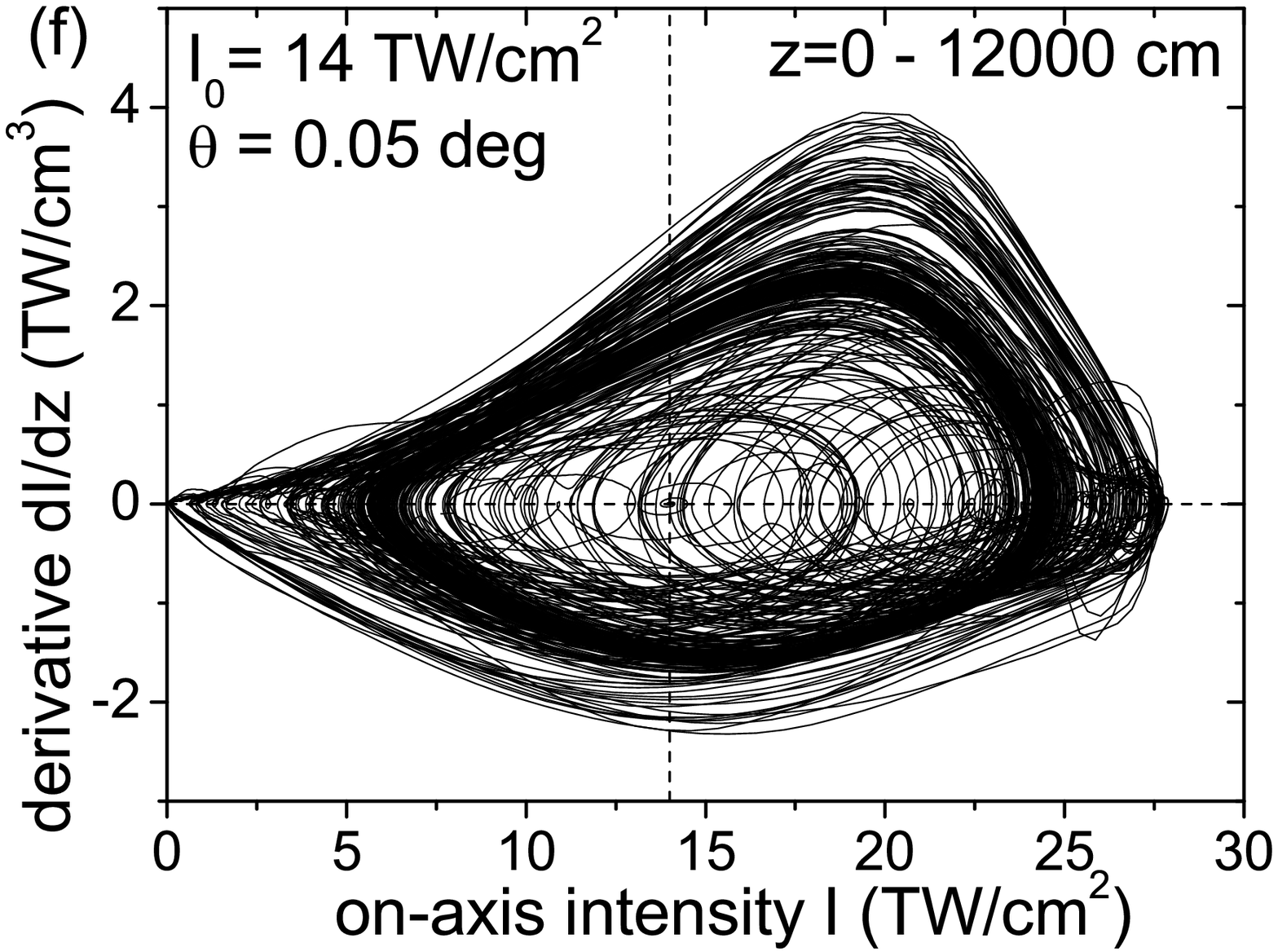}
\end{center}
\caption{\label{Fig6} In air at 800 nm, on-axis intensity $I$ versus propagation distance $z$ of perturbed, ideal NL-UBBs of peak intensities intensities $I_0=14$ TW/cm$^2$ and cone angles (a) $\theta=0.075$ deg and (b) $\theta=0.05$ deg. (c) and (d) Corresponding phase spaces $I$--$dI/dz$ in $[400,3000]$ cm (beginning of the large perturbation regime). (e) and (f) Corresponding phase spaces for the whole propagation range $[0,12000]$ cm. The dashed lines locate the attractor.}
\end{figure*}

For fundamental (vortex-less) NL-UBBs, radial instability appears to be the dominant instability, since no azimuthal breaking has been observed in experiments and simulations \cite{PORRAS1,POLESANAPRE2006,GAIZAUKAS,POLESANAPRA2008,COUAIRON}, particularly under soft input conditions \cite{POLESANAPRL2007}. Linearized stability analysis of NL-UBB against radial perturbations has been performed numerically for typical values of NL-UBB parameters in Ref. \cite{PORRAS1} and \cite{POLESANAPRL2007}, where all details of the procedure are explained. In short, supposing a solution to the NLSE (\ref{NLSE}) of the form
\begin{figure*}
\begin{center}
\includegraphics[width=5.3cm]{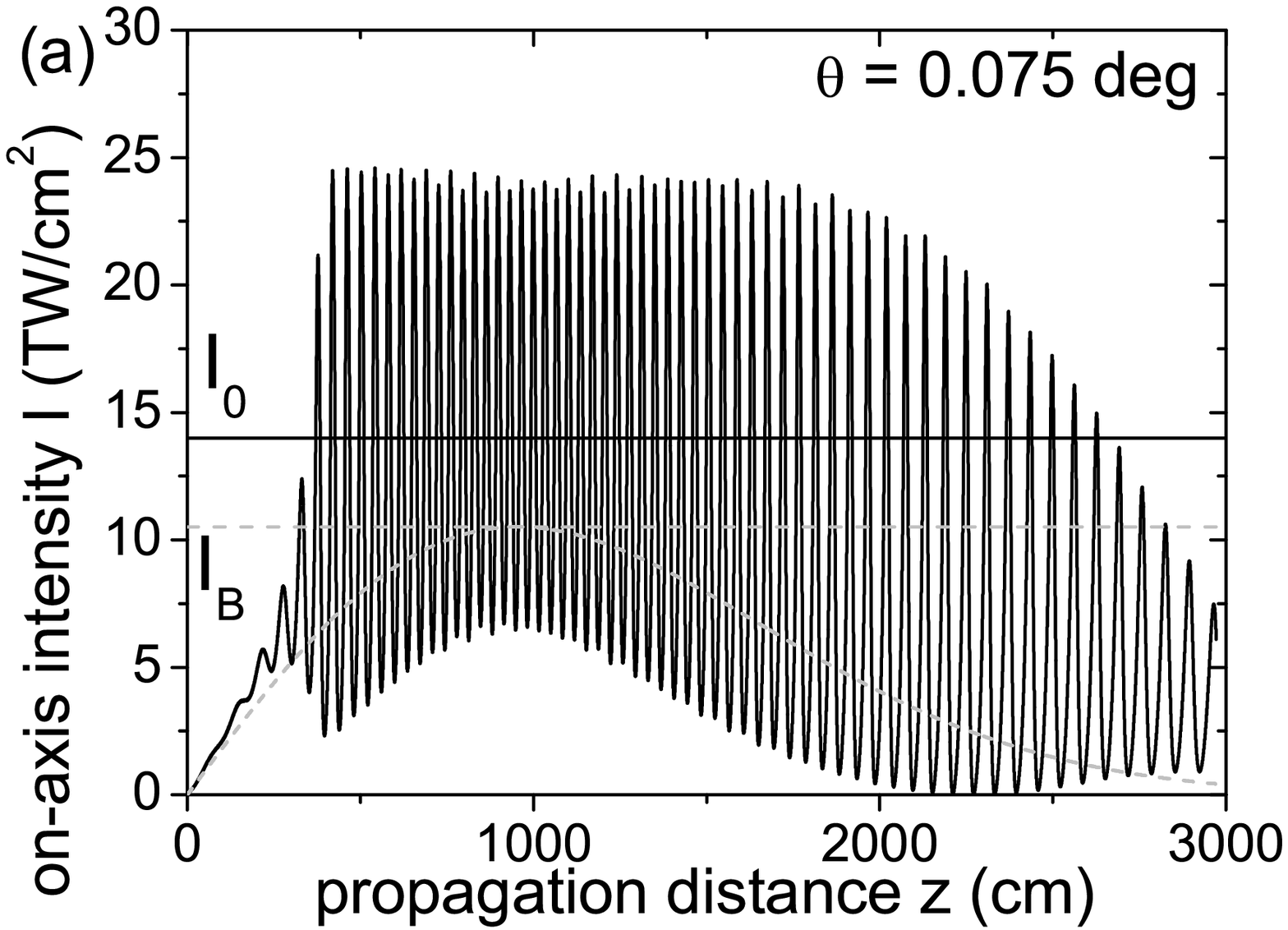}\includegraphics[width=5cm]{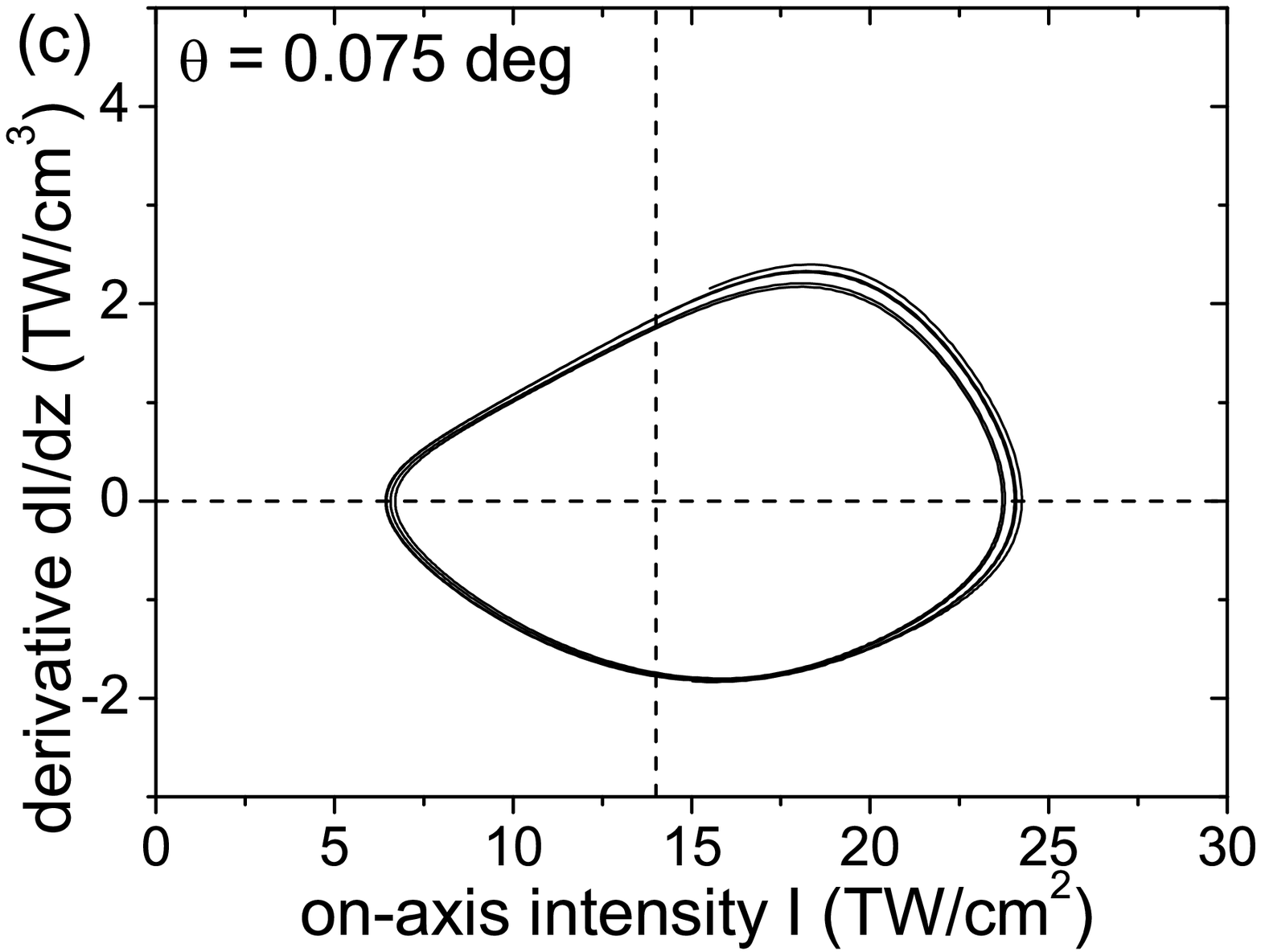}
\includegraphics[width=5.3cm]{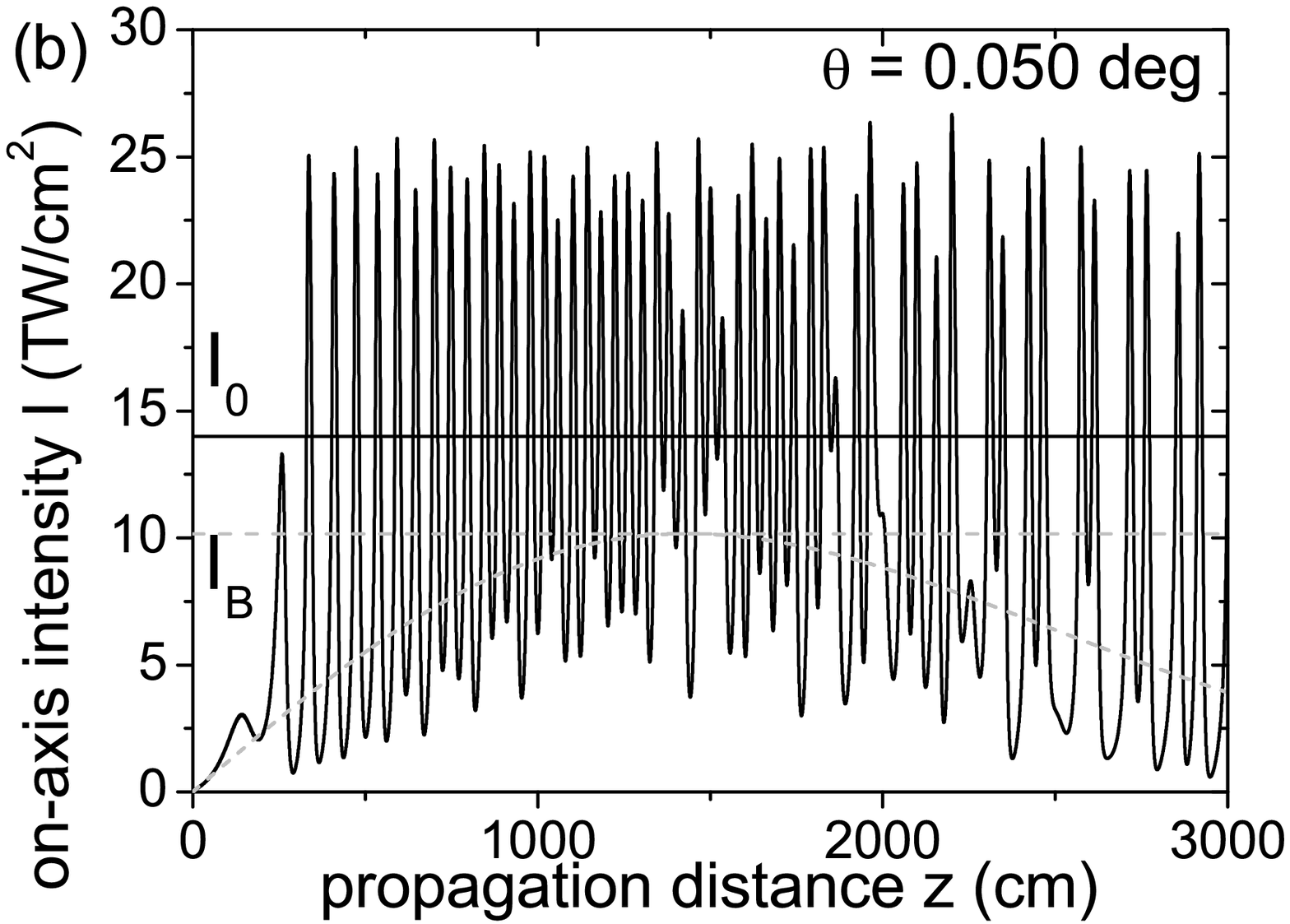}\includegraphics[width=5cm]{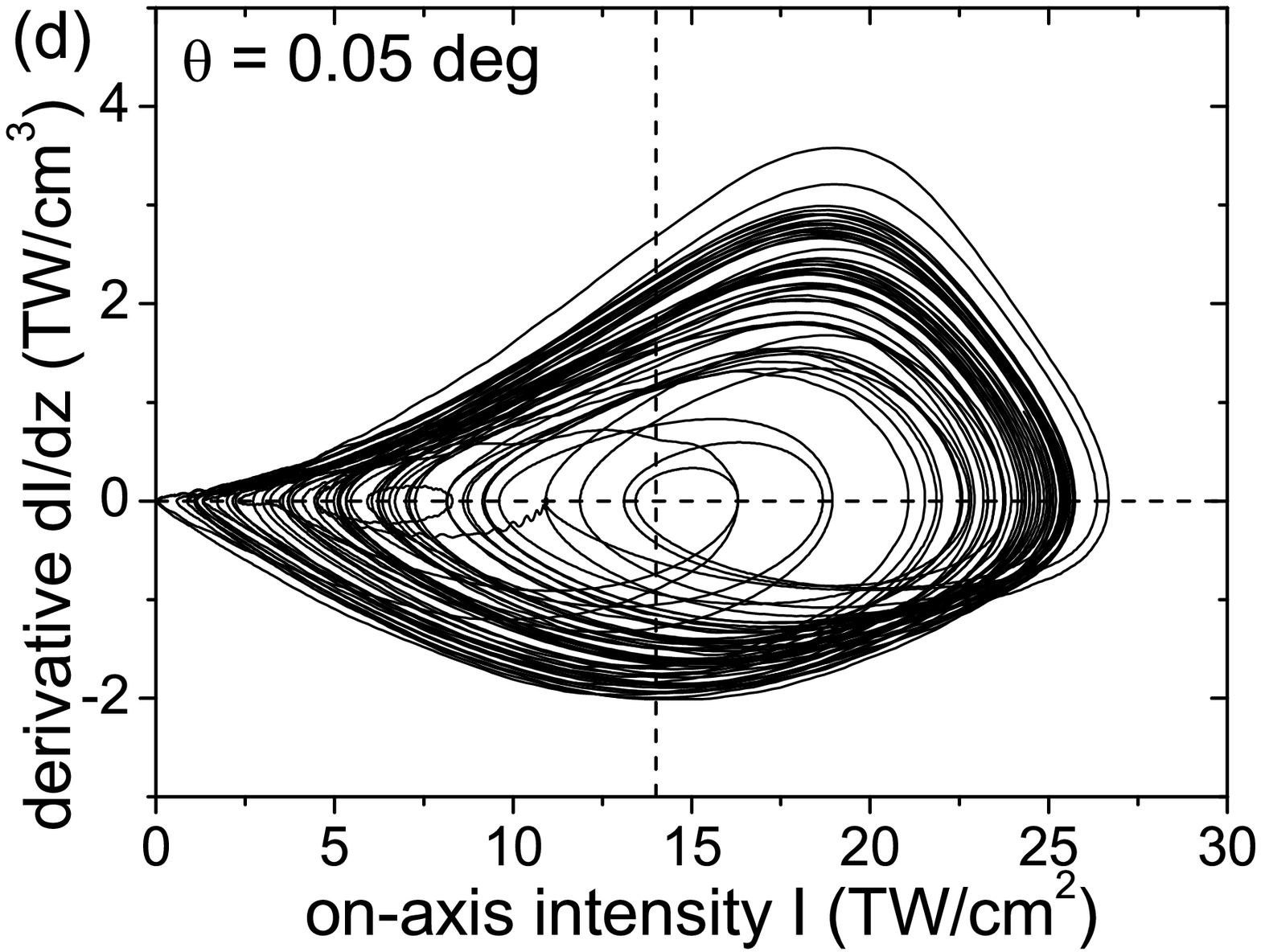}
\end{center}
\caption{\label{Fig7} (a) On axis intensity $I$ (solid curve) in air at 800 nm after and axicon imprinting a cone angle $\theta=0.075$ deg illuminated by a Gaussian beam of width $w=2.5$ cm and intensity $I_G=0.02146$ TW/cm$^2$. In linear propagation (dashed curve) the intensity of the BB would be $I_B=10.508$ TW/cm$^2$ (horizontal dashed line) at $z_B=955$ cm, so that the attracting NL-UBB is defined by $I_0=14$ TW/cm$^2$ (horizontal solid line) and $\theta=0.075$ deg. (b) The same except that $\theta=0.05$ deg and $I_G=0.03113$ TW/cm$^2$. In linear propagation the BB of intensity $I_B=10.165$ TW/cm$^2$ would be formed at $z_B=1432$ cm, so that the attracting NL-UBB is defined by $I_0=14$ TW/cm$^2$ and $\theta=0.05$ deg. (c) Phase space $I$ -- $dI/dz$ about the focal region of the axicon in case (a). (d) Phase space in the whole propagation in case (b). The dashed lines in (c) and (d) locate the attractor.}
\end{figure*}
\begin{equation}\label{PER}
A = a(r)e^{i\phi(r)}e^{i\delta z} + \epsilon[u(r)e^{i\kappa z} + v^\star(r)e^{-i\kappa^\star z}]e^{i\delta z}\,,
\end{equation}
that is, a NL-UBB plus a small ($\epsilon\rightarrow 0$) mode $(u,v)$ that grow exponentially in case that $\mbox{Im}\kappa <0$ while (possibly) oscillating harmonically with frequency $\mbox{Re}\kappa$, a differential eigenvalue problem is obtained for the eigenvalues $\kappa$ and eigenmodes modes $(u,v)$, which has to be solved numerically. As pointed out in \cite{POLESANAPRL2007}, the difficulty with this analysis for NL-UBBs compared to that for standard solitons lies in the weak localization of NL-UBBs. Truncation of the NL-UBB in any finite radial box imposed by the numerical procedure sets a lower bound to the reliable values of $|\mbox{Im}\kappa|$ \cite{POLESANAPRL2007}. Figure \ref{Fig5} shows examples of the exponential gain $-\mbox{Im}\kappa$ and the oscillation frequency $\mbox{Re}\kappa$ of the most unstable mode of NL-UBBs in air at 800 nm as functions of the cone angle and fixed values of the peak intensity [Figs. \ref{Fig5}(a) and (b)], and as functions of intensity and fixed values of the cone angle [Figs. \ref{Fig5}(c) and (d)]. As noted in \cite{POLESANAPRA2008}, NL-UBBs tend to stabilize as the cone angle increases. According to our analysis, no signs of instability are present for $\theta$ above a certain threshold angle (about $\theta\simeq 0.23$ deg in Fig. \ref{Fig5}), but a definitive response to the question of the absolute stabilization of NL-UBBs cannot be given. The trend of $-\mbox{Im}\kappa$ with increasing cone angle suggests an exponential decay. With increasing intensity, Kerr nonlinearity renders NL-UBBs increasingly unstable at first, but the increasing NLLs has an opposite stabilizing effect at higher intensities. Stabilization by NLL appears to be complete above a certain cone angle (about $\theta\simeq 0.15$ deg in Fig. \ref{Fig5}). It is interesting that below this angle, once the Kerr-induced unstable mode disappears by the action of NLLs (at about $I_0=18$ TW/cm$^2$), an underlying unstable mode with a different oscillation frequency $\mbox{Re}\kappa$ becomes dominant.

In connection with Figs. \ref{Fig2}(a) and \ref{Fig3}(a), the steady propagation regime after the axicon can be seen to be associated with the absence of unstable modes of the attracting NL-UBB with $\theta=0.15$ and $I_0=28$ TW/cm$^2$ that tends to be formed at the center of the Bessel zone. Even if repeated self-focusing cycles are observed before the focus of the axicon for these small cone angles (cycles that may be much more pronounced), and  before the waist of the Bessel-Gauss beam, the input radiation is pushed stably towards the NL-UBB about the center of the Bessel zone. In Fig. \ref{Fig2}(c) and Fig. \ref{Fig3}(b), instead, the (weak) unstable regime appears to reflect the instability of the attracting NL-UBB with $\theta=0.15$ and $I_0=12$ TW/cm$^2$. The gain is indeed low [Fig. \ref{Fig5}(c)] (compared to next situations below), and the oscillation frequency about the focus or waist is seen to coincide with the oscillation frequency $\mbox{Re}\kappa$ of the dominant unstable mode of the attracting NL-UBB [Fig. \ref{Fig5}(d)].

The connection between unsteady Bessel propagation regime and instability of the attracting NL-UBB is clearer in situations of higher gain.
For two NL-UBBs with increasing gain, Fig. \ref{Fig6}(a) and (b) shows the growth of the respective dominant unstable modes with propagation distance. These small modes develop into large, periodic (but no longer harmonic) perturbation regimes, that gradually turn into quasi-periodic, and eventually into chaos. In all cases we have studied, this process is found to be faster as the gain $-\mbox{Im}\kappa$ triggering this process is higher. Also, the oscillation frequency in the large, periodic perturbation regime is close to but slightly lower than $\mbox{Re}\kappa$. In simulations as those of Fig. \ref{Fig6}, NL-UBBs are directly launched into the medium and the dominant, unstable modes of each NL-UBB are found to emerge spontaneously from numerical noise with the gain and oscillation frequency predicted by the linearized instability analysis [left part in Figs. \ref{Fig6}(a) and (b)]. To simulate the propagation of ideal, non-truncated NL-UBBs, we use the procedure of replacing the propagated field at each axial numerical step of propagation with the initial NL-UBB in a narrow annulus touching the end of the (quite large) numerical radial grid. This procedure is justified since no dynamics is expected to take place in the linear asymptotic tails. In Figs. \ref{Fig6}(c-f), phase spaces $I$--$dI/dz$ ($I$ on-axis intensity) in relevant propagation intervals are shown. In the case of lower gain, the large perturbation regime remains periodic for a considerable propagation distance [Fig. \ref{Fig6}(c)], whereas in the case of higher gain, it becomes quasi-periodic from the beginning of the large perturbation regime [Fig. \ref{Fig6}(d)] and enters sooner into chaos. The phase spaces up to the longest propagation distance [Figs. \ref{Fig6}(e) and (f)] evidence that NL-UBBs are actually chaotic attractors, whose morphology depends on the specific NL-UBB.

On the other hand, Figs. \ref{Fig7}(a) and (b) show the on-axis intensities after an axicon illuminated with two Gaussian beams of width and intensities such that the attractors are the two NL-UBBs analyzed above. In the case of Fig. \ref{Fig7}(a) with lower gain, the oscillation frequency about the focus of the axicon coincides with that of the large, periodic, perturbation regime. Indeed the structure of the oscillations in the phase space of Fig. \ref{Fig7}(c) about the focus of the axicon mimics the structure of the anharmonic oscillations in the periodic perturbation regime of the attracting NL-UBB in Fig. \ref{Fig6}(c). Small differences originate from the slow decay of intensity along the Bessel zone of the finite-power BB. The structure of the phase space in the whole Bessel zone (not shown) does not reproduce the morphology of the chaotic attractor in Fig. \ref{Fig6}(e). In the case of Fig. \ref{Fig7}(b) with higher gain, the on-axis intensity in a considerable part of the Bessel zone exhibits a highly disordered dynamics. Comparison of the morphology of the phase space in Fig. \ref{Fig7}(d) for the whole Bessel zone with that of the attracting NL-UBB in Fig. \ref{Fig6}(f) evidences that the dynamics after the axicon is reproducing the chaotic dynamics about the ideal NL-UBB chaotic attractor.

\section{Conclusions}

From a series of diagnostic numerical simulations, we have extracted the underlying laws governing the spatial dynamics of the light beam emerging from an axicon, and entering a medium where self-focusing Kerr effect and multiphoton absorption are relevant. If as pointed out, temporal and plasma effects play a secondary role in determining the spatial dynamics in filamentation with BBs, these laws provide an unified understanding of the different Bessel filamentation regimes described previously. In a few words, the nonlinear propagation is determined by an attracting NL-UBB and its stability properties under small perturbations. The attracting NL-UBB is that whose inward H\"ankel amplitude equals the amplitude of the BB that the BB generator would create at the center of the Bessel zone in linear propagation. We have derived an approximate analytical expression that determines the attracting NL-UBB given the optical properties of the medium, the cone angle, and the intensity of the linear BB (or equivalently, the axicon base angle and the input Gaussian width and intensity). Steady/unsteady propagation regimes are shown to correspond to stability/instability of the attracting NL-UBB under small radial perturbations, i. e., to the existence of a small unstable radial mode that tends to grow exponentially. We have performed an extensive stability analysis under small radial perturbations that put in quantitative terms the stabilization effect of increasing the cone angle and the intensity. In case of instability under small perturbations, NL-UBBs are seen to develop a large perturbation and chaotic regimes with increasing propagation distances. In the Bessel zone after the axicon, and depending on how large the gain of the small unstable mode of the attracting NL-UBB is, the unsteady dynamics reproduces the dynamics of the small perturbation, large, or chaotic perturbation regimes of the attracting NL-UBB. Though a direct relation with increasing gain is obvious, further research would be needed to specify in more quantitative terms the particular perturbation regime (small, large or chaotic) of the attracting NL-UBB that is observed in the Bessel zone. We have restricted ourselves to vortex-less NL-UBBs, but the generality of these ideas suggests a relatively simple generalization to axicon-generated vortex NL-UBB, in which case not only radial instability but also azimuthal instability should be taken into account.

\acknowledgments

M.A.P. acknowledges support from Projects of the Spanish Ministerio de Econom\'{\i}a y Competitividad
No. MTM2012-39101-C02-01 and  No. FIS2013-41709-P. J.C.L. acknowledges support from Project of the Spanish Ministerio de Econom\'{\i}a y Competitividad MTM2012-39101-C02-01.

\end{document}